\documentclass[prd,twocolumn,reprint,preprintnumbers,nofootinbib]{revtex4-1}

\input{header.tex}

\begin{document}

\reportnum{-2}{CERN-TH-2024-157}

\title{Searches for long-lived dark photons at proton accelerator experiments}
\author{Yehor Kyselov}
\email{kiselev883@gmail.com}
\affiliation{Taras Shevchenko National University of Kyiv, Kyiv, Ukraine}
\author{Maksym~Ovchynnikov}
\email{maksym.ovchynnikov@cern.ch}
\affiliation{Theoretical Physics Department, CERN, 1211 Geneva 23, Switzerland}

\date{\today}

\begin{abstract}
A systematic and unified analysis of the capability of lifetime frontier experiments to probe the parameter space of hypothetical long-lived particles is crucial for defining the targets of future searches. However, such a comprehensive study has not yet been conducted for dark photons -- hypothetical massive particles that kinetically mix with Standard Model photons. Existing studies often rely on outdated assumptions about dark photon phenomenology, leading to inaccurate calculations of the parameter space, including regions already excluded or relevant for future experiments. In this paper, we present a comprehensive calculation of the parameter space for GeV-scale dark photons and light dark matter accessible to lifetime frontier experiments. Our analysis highlights how theoretical uncertainties in dark photon phenomenology can significantly impact experimental sensitivity across coupling and mass parameters. To facilitate further investigation, we provide these results in a publicly accessible form, incorporating updated dark photon phenomenology into the event generator \texttt{SensCalc}.
\end{abstract}

\maketitle

\section{Introduction}
\label{sec:introduction}

The inability of the Standard Model (SM) of particle physics to explain several observed phenomena -- such as the baryon asymmetry of the Universe, neutrino oscillations, and dark matter -- suggests the potential existence of new particles yet to be discovered. A class of such particles falls within the GeV mass range. If they exist, their detection has eluded us so far due to their very weak interaction couplings. These particles are generally unstable, but their relatively low mass and tiny couplings make them long-lived, with characteristic decay lengths $c\tau \gamma$ significantly exceeding the spatial dimensions of typical laboratory experiments. Because of this, they are often called Long-Lived Particles, or LLPs.

Various interactions of LLPs can be probed in experiments that involve proton beam collisions with other beams or fixed targets. A wide array of such experiments has operated in the past and continues to run today. The growing interest in LLPs has led to numerous proposals over the past decade to explore the lifetime frontier~\cite{Alekhin:2015byh,Beacham:2019nyx,Antel:2023hkf}, with notable examples including the ongoing NA62~\cite{NA62:2023qyn}, SND@LHC~\cite{SNDLHC:2022ihg} and FASER~\cite{FASER:2019aik} experiments, the recently approved DUNE~\cite{DUNE:2020lwj}, SHiP~\cite{Aberle:2839677} and DarkQuest~\cite{Batell:2020vqn,Apyan:2022tsd}, as well as proposed experiments like advSND~\cite{Abbaneo:2895224}, FASER2~\cite{Kling:2021fwx}, FACET~\cite{Cerci:2021nlb}, MATHUSLA~\cite{MATHUSLA:2019qpy}, and ANUBIS~\cite{Bauer:2019vqk}.

Understanding the potential of lifetime frontier experiments to explore LLPs requires addressing a key question: how well do we know the phenomenology of these particles, including how they are produced and how they decay? The answer is far from satisfactory. Each LLP has unique production and decay modes, often involving strong interactions, leading to significant complexities. Specifically, there is no robust description of these processes at transferred momenta on the order of GeV, which corresponds to the mass range of GeV-scale LLPs. Moreover, discrepancies in the way different experiments describe LLP phenomenology have introduced inconsistent variations in reported constraints and sensitivities (see the discussion in~\cite{Ovchynnikov:2023cry}). As a result, summary plots of LLP parameter spaces often show divergent results and require systematic revision. These challenges underscore the importance of a comprehensive study of LLP phenomenology and the need to revisit the sensitivity reach of both past and future experiments.

One of the most studied and well-motivated models illustrating these issues is dark photon -- a hypothetical particle that mixes kinetically with Standard Model photons. The relevant Lagrangian is:
\begin{equation}
\mathcal{L} = \frac{\epsilon}{2}F_{\mu\nu}V^{\mu\nu},
\end{equation}
where $F_{\mu\nu}$ and $V_{\mu\nu} = \partial_{\mu}V_{\nu} - \partial_{\nu}V_{\mu}$ are the field strengths of the SM and dark photons, respectively, $V_{\mu}$ is the dark photon field, and $\epsilon \ll 1$ is the mixing parameter. This model is minimal, renormalizable, and offers a rich physics case. For instance, dark photons, decaying both visibly and invisibly, could mediate interactions between the SM and hypothetical GeV-scale dark matter (see~\cite{Beacham:2019nyx,Battaglieri:2017aum,Kahlhoefer:2015bea,Duerr:2016tmh,Garcia:2024uwf,Fabbrichesi:2020wbt} and references therein). 

The dark photon enters two benchmark models (referred to as BC1 and BC2) proposed by the Physics Beyond Colliders (PBC) initiative to evaluate the reach of future experiments~\cite{Beacham:2019nyx}. Many studies have considered these models to present constraints and sensitivities, see, e.g.,~\cite{deNiverville:2011it,deNiverville:2012ij,Blumlein:2013cua,Batell:2014mga,Gorbunov:2014wqa,deNiverville:2016rqh,MiniBooNEDM:2018cxm,SHiP:2020sos,SHiP:2020vbd,Feng:2022inv,MiniBooNEDM:2018cxm,Kling:2021fwx,Kling:2022ehv,Ovchynnikov:2023cry,Asai:2022zxw,Asai:2023mzl,Chakraborty:2021apc,Barman:2024lxy,Garcia:2024uwf}.

However, there are a few problems preventing us from a comprehensive understanding of the laboratory reach of the parameter space of dark photons. The first problem concerns the maturity of the description of some of its production channels in the literature. For example, proton bremsstrahlung, a major production channel for GeV-scale dark photons, has been primarily modeled using the widely cited approach~\cite{Blumlein:2013cua}, which does not properly account for theoretical uncertainties. In the case of production via mixing with neutral mesons, the dark photon cross-section is often approximated by multiplying the meson production cross-section by the squared mixing angle. This approximation fails to accurately represent both the total yield and the kinematics of the $V$s.

In addition, the description of dark photon phenomenology has been used inconsistently across various experiments. For instance, consider the constraints and sensitivities on the parameter space of dark photons $V$ presented in overview papers~\cite{Beacham:2019nyx,Antel:2023hkf}. Different treatments of the proton elastic form factor, determining the production probability of the bremsstrahlung mechanism, have been used to calculate the constraints. Namely, the form factor used in obtaining the bound from experiments like NA62 and NuCal did not include resonant contributions from dark photon mixing with neutral vector mesons. On the other hand, the rest of the experiments incorporate this effect, although they parametrize this contribution in a non-unique way~\cite{deNiverville:2016rqh,Berlin:2018jbm,SHiP:2020sos}. This is particularly relevant as NA62 has recently published updated world-leading constraints based on their 2021 data collection~\cite{NA62:2023qyn}.

Similar inconsistencies arise in the treatment of hadronic decays, which affect both the dark photon’s lifetime and its visible decay channels. 

In this paper, we address these issues by systematically calculating the reach of accelerator-based experiments for long-lived dark photons in the GeV mass range. We incorporate recent advancements in dark photon phenomenology: Ref.~\cite{Foroughi-Abari:2024xlj}, which reanalyzed proton bremsstrahlung production, and forthcoming work~\cite{Kyselov:2025ta}, which investigates production via mixing within the fragmentation chains in deep inelastic collisions. We then implement this updated phenomenology in \texttt{SensCalc}~\cite{Ovchynnikov:2023cry}, a tool designed to calculate event rates for LLPs across various experimental setups in a unified framework. It accounts for the acceptance of LLP decay products in the detector and applies various selection criteria based on event topology, geometric parameters, and energy. Using this, we ``recalibrate'' the parameter space for dark photons and evaluate the effect of bremsstrahlung uncertainties on the experimental reach.

The paper is organized as follows. In Sec.~\ref{sec:phenomenology}, we provide a comprehensive discussion of dark photon phenomenology, including production and decay modes, and parameterize the uncertainties in proton bremsstrahlung. Sec.~\ref{sec:method} describes the methods used to calculate the experimental reach, including the implementation in \texttt{SensCalc}. In Sec.~\ref{sec:revisiting}, we present recalculated constraints and sensitivities for various experiments, showing how the revised production channels, particularly uncertainties in bremsstrahlung, affect these results. There, we also consider the case of the elastic dark matter coupled to dark photons -- another widely considered model of new physics. Finally, Sec.~\ref{sec:conclusions} contains our conclusions.

\section{Dark photon phenomenology at proton accelerators: overview and implementation}
\label{sec:phenomenology}

\subsection{Production modes}
\label{sec:production}

\begin{figure*}
    \centering
    \includegraphics[width=0.9\textwidth]{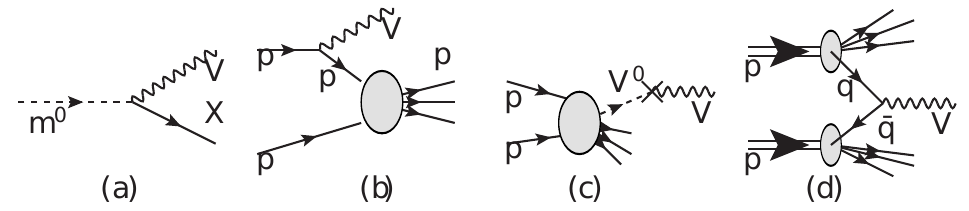}
    \caption{The diagrams of various production processes of dark photons $V$: decays of neutral mesons $m^{0}\to V+X$ (the diagram (a)), proton bremsstrahlung (the diagram (b)), the production via mixing with neutral mesons $V^{0}$ (the diagram (c)), and the Drell-Yan process (the diagram (d)).}
    \label{fig:production-diagrams-vector}
\end{figure*}

Dark photons may be viewed as photons with non-zero mass. Their production modes (see Fig.~\ref{fig:production-diagrams-vector}) may be, therefore, understood through the prism of this logic:
\begin{itemize}
\item[--] Decays of pseudoscalar mesons $m^{0} = \pi^{0},\eta,\eta'$, mainly the anomalous processes $m^{0}\to \gamma V$ (similarly to the SM process $m^{0}\to \gamma \gamma$).
\item[--] The proton bremsstrahlung, where dark photons get elastically emitted from the proton leg, with the rest being the inelastic scattering.
\item[--] The Drell-Yan process, which is a fusion of two partons from the colliding protons~\cite{SHiP:2020vbd,Kling:2021fwx,Ovchynnikov:2023cry}.
\item[--] The production via mixing with vector mesons $V^{0} = \rho^{0}, \omega, \phi$ and their excitations that appear during the fragmentation of partons.
\item[--] Cascade production of dark photons in the case of beam dump experiments. It originates from the interactions of secondary particles inside the thick target. Besides enhancing the fluxes of $\pi^{0}$, it also introduces the production from electron-positron annihilation and $e^{\pm}$ bremsstrahlung~\cite{Blinov:2024pza}.
\end{itemize}

For meson decays and the Drell-Yan process, we adopt the standard framework from the literature~\cite{SHiP:2020sos, Ovchynnikov:2023cry}. Specifically, for meson production, we simulate the sequential generation of light mesons using \texttt{PYTHIA8}~\cite{Bierlich:2022pfr}, followed by their decays into dark photons. In the case of LHC-based experiments relevant to dark photons, we configure \texttt{PYTHIA8} with the forward tune~\cite{Fieg:2023kld}, while for lower-energy experiments, such as those at SPS, Fermilab, and Serpukhov, we use the setup described in~\cite{Dobrich:2019dxc}. For the Drell-Yan process, we generate the leading-order parton-level interaction $q+\bar{q} \to V$ in \texttt{MadGraph}~\cite{Alwall:2014hca} and apply showering and hadronization using \texttt{PYTHIA8} to capture the transverse momentum distribution. By default, we use the parton distribution function (PDF) set \texttt{NNPDF 3.1 NNLO}, which is typically utilized for the calculations with new physics particles~\cite{SHiP:2020vbd, Berlin:2018jbm}. We estimate uncertainties by varying the renormalization and factorization scales around their central values by a factor of two (similar to Refs.~\cite{SHiP:2020vbd, Berlin:2018jbm}), as well as varying the PDFs themselves, treating the uncertainties from the former and the latter as statistically independent. Typically, these uncertainties remain within a factor of a few but increase at small dark photon masses, $\mathcal{O}(1\text{ GeV})$. The latter is caused by peaking the production of light particles in the quark fusion in the region of small Parton's energy fraction $x$, which is one of the least experimentally explored domains in PDFs. This feature is especially important for experiments with high collision energies, such as LHC.

Regarding the production via mixing, we implement this mechanism directly in \texttt{PYTHIA8}~\cite{Bierlich:2022pfr} by replacing the intermediate mesons $V^{0}$ with dark photons at the final stage of the fragmentation chain. This approach allows for a systematic calculation of the total yield and kinematics of long-lived particles (LLPs) as a function of their mass. Such a feature was absent in previous studies~\cite{Berlin:2018jbm,Jerhot:2022chi}, which approximated the LLP flux by multiplying the meson production cross-section with the squared mixing angle and either assumed identical kinematics for the LLPs and their parent mesons or used unclear relationships.

More details on this implementation will be provided in a forthcoming paper~\cite{Kyselov:2025ta}. The primary uncertainty in this channel stems from variations in the \texttt{PYTHIA} setup, which needs to be tuned to match accelerator data. For the far-forward LHC experiments, we adopted the tune from~\cite{Fieg:2023kld}, finding that the flux varies by about 30\%, largely independent of the LLP mass. Detailed tuning studies for other facilities are currently lacking and are deferred to future work. However, following the LHC case, we do not expect the variation of the uncertainty within the range beyond $\mathcal{O}(1)$. Also, we do not include the contribution from higher excitations of vector mesons, which may sizeably enhance the flux from mixing at masses $m_{V}\gtrsim 1\text{ GeV}$. The reason is that the production of such mesons in the fragmentation chain is not incorporated in \texttt{PYTHIA8}. Nevertheless, the resulting flux is a conservative estimate. 

Let us switch to the proton bremsstrahlung.\footnote{A detailed description of this production mechanism is provided in Appendix~\ref{app:bremsstrahlung}.} Consider the cross-section of the scattering of two protons $P,p$ where dark photons are emitted from the proton leg. If concentrating solely on the initial state radiation part of the process, its cross-section may be approximated by the cross-section of the inelastic scattering of two protons $p,p'$ times the ``splitting probability'' $\omega_{\text{spl}}$ describing the virtual transition $P\to p'+V$. Another key quantity that defines the bremsstrahlung flux is the elastic proton form-factor. Because of the mixing of $V$ with vector mesons $V^{0}$, it gets resonantly enhanced at the vicinity of $V^{0}$ masses.  

Previous studies of laboratory reach to the models with dark photons in the GeV mass range have predominantly used the description of $\omega_{\text{spl}}$ from~\cite{Blumlein:2013cua}. This study computes the differential cross-section for quasi-elastic bremsstrahlung using a variant of the Weizsäcker-Williams approximation, subsequently extrapolating this to cover the full inelastic process. This definition of $\omega_{\text{spl}}$ is accompanied by the description of the proton elastic form factor from~\cite{Berlin:2018jbm}, which uses the minimal vector meson dominance model and includes the three lowest excitations of $\rho$ and $\omega$.

However, the approach described above does not properly quantify the theoretical uncertainties arising from introducing the splitting probability. It leaves ambiguities in defining key kinematic distributions of dark photons (such as the typical transverse momenta $p_T$) and the overall bremsstrahlung flux.

Ref.~\cite{Foroughi-Abari:2024xlj} considered another approach based on the idea of the parton splitting kernel from~\cite{Altarelli:1977zs} (see also earlier studies~\cite{Boiarska:2019jym,Foroughi-Abari:2021zbm,Gorbunov:2023jnx}). Its transparency allows for clearly identifying the main theoretical uncertainty -- the virtuality of $p'$. An independent source for the uncertainty (also studied in~\cite{Foroughi-Abari:2024xlj}) is the elastic $ppV$ form factor. Ref.~\cite{Foroughi-Abari:2024xlj} has added excitations of $\phi$ and considered two models -- the dispersion relation method and the unitary and analytic model -- to calculate the combined contributions of vector mesons $V^{0}$. The uncertainty has been estimated by varying the parameters entering the analytic expression of the form-factor -- masses and widths of $V^{0}$s and their excitations. 

Being combined, the two uncertainties lead to large variations in the production flux, potentially spanning orders of magnitude, and must, therefore, be carefully accounted for in the dark photon parameter space.

To reflect the overall bremsstrahlung uncertainty and compare different descriptions to describe this production channel, we consider the following descriptions:
\begin{itemize}
    \item[--] The ``Baseline'' setup commonly used in the previous studies. For the splitting probability $\omega_{\text{spl}}$, it assumes the calculation from~\cite{Blumlein:2013cua}, together with the following kinematic cuts: $p_{T} < p_{T,\text{max}} =  1\text{ GeV}$, $\text{min}[0.1,100\text{ GeV}/E_{p}]<z<0.9$, where $z = E_{V}/E_{p}$ is the fraction of the incoming proton energy carried by $V$. The proton elastic form factor is taken from~\cite{Berlin:2018jbm}. 
    \item[--] The ``FR'' method, assuming the splitting probability and the proton inelastic form factor from~\cite{Foroughi-Abari:2024xlj}. To define the uncertainty band, we will vary the hard scale in the virtuality form-factor with $\Lambda_{p}$, as well as consider the variations of the elastic proton form-factors $F_{1},F_{2}$ within the Unitary Analytic model (see also~\cite{Dubnicka:2002yp}), which is a conservative choice.
\end{itemize}
Within the FR description, we define three characteristic setups: \texttt{central}, which corresponds to $\Lambda_{p} = 1 \text{ GeV}$ and central values of the form-factors $F_{1,2}$; \texttt{lower}, corresponding to $\Lambda_{p} = 0.5\text{ GeV}$ and lower boundary of the form-factors uncertainty; and \texttt{upper}, for which $\Lambda_{p} = 2\text{ GeV}$ and the form-factors are fixed by the upper boundary. 

Ref.~\cite{Foroughi-Abari:2024xlj} considered the variation of $\Lambda_{p}$ within a somewhat narrower range $1-2\text{ GeV}$, motivating this by the fact that it does not lead to the overproduction of $\rho^{0}$ mesons compared to the data coming from the NA27 experiment. However, the comparison was not apple-to-apple: the real data is the inclusive production of $\rho^{0}$ mesons that include both initial and final state radiation processes, as well as not-so-negligible secondary production originating from the finite width of the target. Because of this, we believe that while the upper bound $\Lambda_{p}= 2\text{ GeV}$ is meaningful, the lower bound may be overestimated. As there is no other indication for this boundary, we are forced to assume a reference choice, and hence choose $0.5$ GeV.

Lastly, for SPS and other beam dump experiments, it may be important to consider the cascade production of dark photons. We do not incorporate this in our analysis. We motivate this for two reasons. First, the secondarily produced $\pi^{0},\eta,\dots$ are much softer than the primary ones, and have a broad angular distribution. These two features significantly decrease their contribution to the flux of $V$s, pointing to the decay volume and having large enough energy to be detected (see a discussion in~\cite{SHiP:2020sos,SHiP:2020vbd}). Second, the $e^{+}e^{-}$ annihilation and $e^{\pm}$ bremsstrahlung mainly enhance the low-mass production flux $m_{V}\lesssim 100\text{ MeV}$ in the domain of relatively low energies. The corresponding parameter space is mainly excluded by astrophysical constraints and past experiments.

\subsubsection{Production channels hierarchy}
\begin{figure}[h!]
    \centering
    \includegraphics[width=0.45\textwidth]{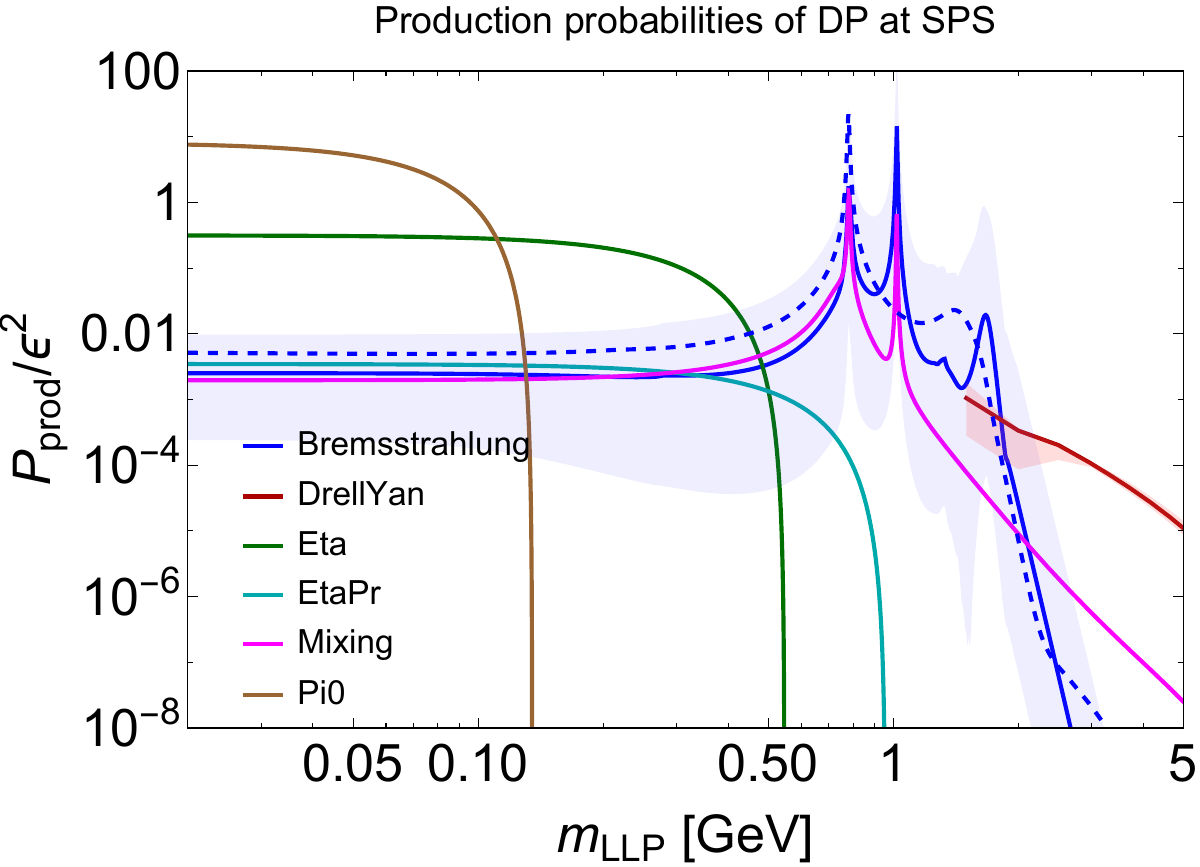} \\ \includegraphics[width=0.45\textwidth]{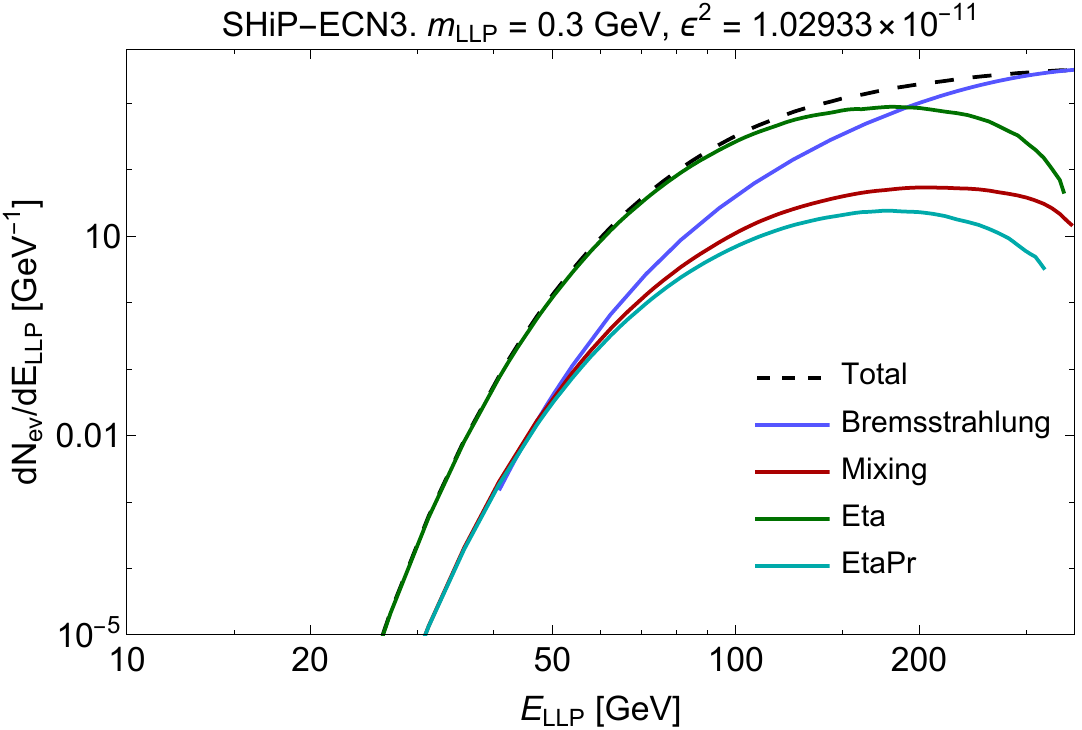}
    \caption{\textit{Top panel}: overall probabilities to produce dark photons (total number of produced dark photons normalized by the number of proton collisions) via various channels described in Sec.~\ref{sec:production}. The SPS facility is considered as an example. The blue curves indicate various descriptions of the bremsstrahlung channel: the Baseline description commonly used in the literature (the dashed blue line) and the FR description from the paper~\cite{Foroughi-Abari:2024xlj} (the blue line), with the uncertainties for the latter indicated by the light blue band. The width of the band is caused by the variation of the proton virtuality scale and the proton elastic electromagnetic form factor in the time-like region (see also Appendix~\ref{app:bremsstrahlung}). The red domain originates from the uncertainty in the description of the Drell-Yan process. \textit{Bottom panel}: the contribution of these production channels to the differential number of events in dark photon's energy at the SHiP experiment. For the bremsstrahlung, the ``Central'' description is used. The value of the mixing angle close to the upper bound of the sensitivity is considered (i.e., $c\tau \langle\gamma \rangle$ is smaller than the distance from the target to the decay volume), which highlights the importance of the bremsstrahlung channel even in the mass range where the overall production probability is suppressed. See Sec.~\ref{sec:method} on the method we used to obtain the distribution.}
    \label{fig:bremsstrahlung-production-probabilities}
\end{figure}

The proton bremsstrahlung, Drell-Yan process, and the production via fragmentation may occur per each generic inelastic proton-proton scattering. In general, there may be an overlap between them. We believe that our approach allows disentangling of these channels, to avoid double counting. Namely, within the approach used to calculate the bremsstrahlung flux, this channel represents the initial state radiation process. To ensure the absence of overlapping with the mixing mechanism, we turn off the initial state radiation in \texttt{PYTHIA8} run. The Drell-Yan process is a specific final state radiation process where dark photons directly appear from quarks (which makes it different from the production via mixing), and it also becomes relevant in the parameter space where mixing and bremsstrahlung become highly suppressed.

The production probabilities for these processes in the particular case of the SPS beam dump facility, with the center-of-mass energy $\sqrt{s}\approx 28\text{ GeV}$, are shown in Fig.~\ref{fig:bremsstrahlung-production-probabilities}. For all the other facilities of interest, the hierarchy between these processes is very similar. Overall, decays of mesons dominate the production in the mass range $m_{V}\lesssim m_{\eta}$, thanks to the large yield of $\pi^{0},\eta$ and parametrically large branching ratios of the decay. The bremsstrahlung production dominates at masses $m_{V}\lesssim 1-2\text{ GeV}$. The mixing channel may become important depending on the position of the bremsstrahlung within the uncertainty band. At higher masses, the Drell-Yan process becomes the dominant one. 

The relative importance of various processes may change depending on the location of the experiment and the LLP lifetimes of interest. For instance, the proton bremsstrahlung largely contributes to the flux of dark photons with tiny polar angles and large energies close to the proton beam energy. Therefore, it is most relevant for on-axis experiments. It may also determine the upper bound of the sensitivity, as the flux of $V$s from decays of $\pi$ and $\eta$ mesons is less energetic. This is illustrated by the lower panel of Fig.~\ref{fig:bremsstrahlung-production-probabilities}.

\subsection{Decay modes}
\label{sec:decays}

\begin{figure}[h!]
    \centering
    \includegraphics[width=\linewidth]{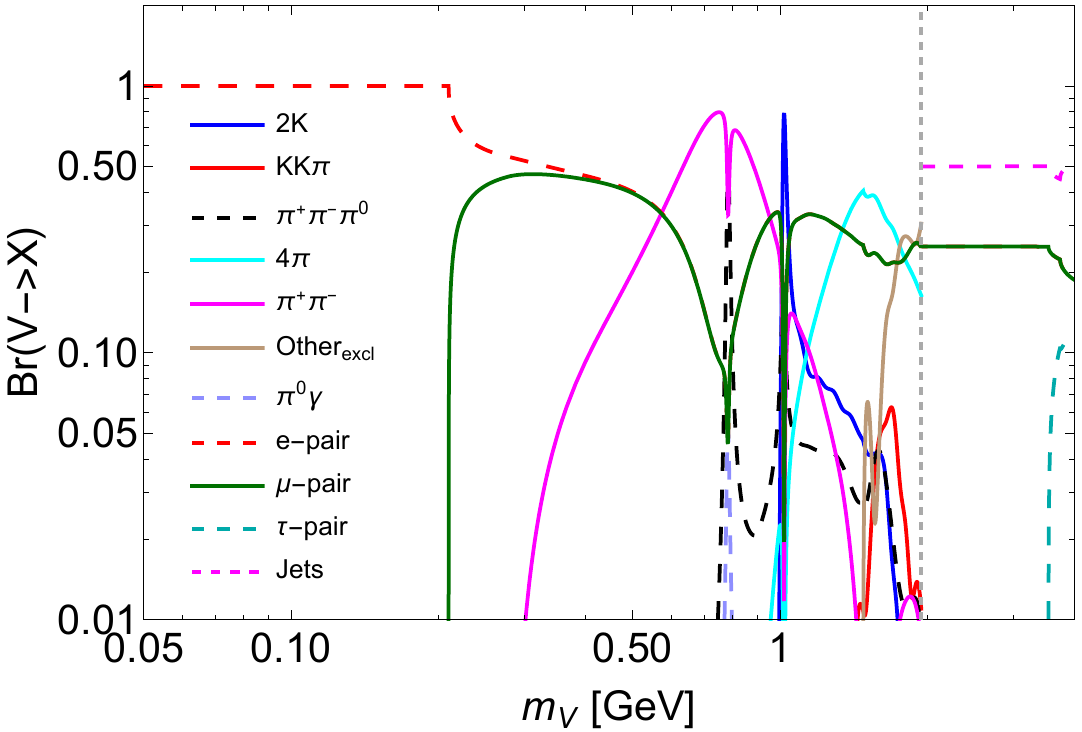} \\ \includegraphics[width=\linewidth]{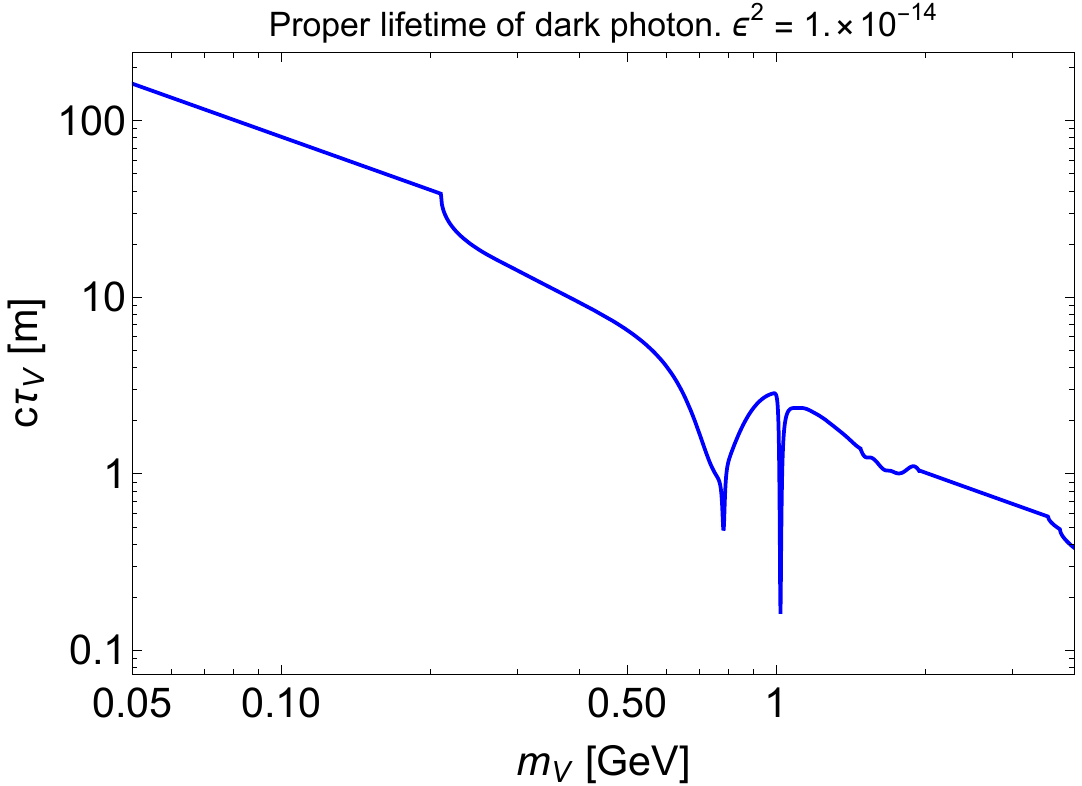}
    \caption{Decays of dark photons. \textit{Top panel}: branching ratios of various decays. To simplify the plot, some channels actually include a few modes. Namely, $2K$ is a sum of $K^{+}K^{-}$ and $K_{L}K_{S}$; $KK\pi$ represents the total probability of the processes $K^{+}K^{-}\pi^{0}$, $K_{S}K^{+}\pi^{-}$ and $K_{S}K^{-}\pi^{+}$; $4\pi$ us the sum of $2\pi^{+}2\pi^{-}$ and $\pi^{+}\pi^{-}2\pi^{0}$; Jets is the collective process $V\to q\bar{q}$, with $q = u,d,s,c$. The resonant features around the masses $\approx 0.78\text{ GeV}$ and $1.02\text{ GeV}$ come from the mixing of dark photons with the $\rho^{0},\omega,\phi$ mesons, that enhances the $\rho/\omega/\phi$-like decay modes. \textit{Bottom panel}: the proper lifetime $c\tau_{V}$. See text for details.}
    \label{fig:decay-br-ratios}
\end{figure}

Decay channels of dark photons include purely leptonic and hadronic processes. 

Leptonic decays are two-body, $V\to l^{+}l^{-}$. Hadronic decays are more subtle. In the limit $m_{V}\gg \Lambda_{\text{QCD}}$, they may be described inclusively -- via decays into a quark-antiquark pair. In the opposite case $m_{V}\lesssim 1\text{ GeV}$, perturbative QCD is inapplicable, and instead exclusive decay modes must be explicitly considered. The information about them may be extracted from the electromagnetic data of the process $e^{+}e^{-}\to \text{hadrons}$ if combining it with the vector meson dominance within the hidden local symmetry approach~\cite{Ilten:2018crw} (see also~\cite{Foguel:2022ppx}). In particular, the squared matrix elements, which are essential to compute the phase space of the decay products, may be obtained from the theoretical approaches similar to~\cite{Fujiwara:1984mp}.

The branching ratios of various decays, as well as the resulting proper lifetime of dark photons, $c\tau_{V}$, are shown in Fig.~\ref{fig:decay-br-ratios}. Leptonic modes dominate in the mass range $m_{V}\lesssim 0.5\text{ GeV}$. The hadronic decays have complicated scaling with mass because of the mixing with $\rho^{0},\omega,\phi$. It leads to the resonant enhancement of the probabilities of the corresponding decays, such as $V\to \pi^{+}\pi^{-}$ (the $\rho^{0}$-like decay), $V\to 2K$ (the $\phi$-like decay), and $V\to 3\pi$ (the $\omega$-like decay).

The inclusive and exclusive descriptions match at mass $m_{V}\approx 1.6\text{ GeV}$. Overall, the theoretical uncertainty coming from this description of dark photon decays is around $20\%$ in the given mass range~\cite{Ilten:2018crw}.

\section{Approach to calculate the reach of experiments}
\label{sec:method}

Having described the dark photon phenomenology, we are now ready to calculate the reach of various experiments to probe its parameter space. 

The main experiments to search for long-lived dark photons in the GeV mass range are proton accelerator experiments. Next, all the production modes are characterized by small transverse momenta, $p_{T} \lesssim 1\text{ GeV}$, meaning that the flux of dark photons is highly collimated along the direction of the incoming proton beam. Given that typical laboratory experiments operate at the facilities with proton center-of-mass energies well above $1\text{ GeV}$, the flux of dark photons is maximized at small polar angles $\theta$. Consequently, experiments in the forward and far-forward directions, especially those located on-beam-axis, are most relevant for these searches.

The experiments we consider are: the past NuCal~\cite{Blumlein:1990ay}, BEBC~\cite{BEBCWA66:1986err,WA66:1985mfx}, and CHARM~\cite{CHARM:1983ayi,CHARM:1985anb}; the currently running NA62~\cite{NA62:2023qyn,Antel:2023hkf}; the approved SHiP~\cite{Alekhin:2015byh,Albanese:2878604} and DarkQuest~\cite{Batell:2020vqn,Apyan:2022tsd}; and the proposed FASER2~\cite{Kling:2021fwx} in its FPF configuration (FASER2-FPF).

In this analysis, we do not consider some of the currently running and future experiments, including FASER, SND@LHC, FACET~\cite{Cerci:2021nlb}, the Downstream algorithm at LHCb~\cite{Gorkavenko:2023nbk}, and DUNE. Despite being the on-axis experiments, the sensitivity of FASER and SND@LHC is constrained by the parameter space where meson decay dominates -- a well-understood regime. This limitation is due to their relatively small geometric acceptance and low integrated luminosity. As such, these experiments do not warrant further revision for this particular study. As for FACET, the Downstream algorithm, and DUNE, their sensitivities may be obtained in a way similar to the sensitivities of the other future experiments (see the section below), and we do not include them in order not to overfill the final plots.

Also, we do not recalculate the constraints from astrophysical probes from supernova, and prompt searches from LHCb~\cite{Ilten:2016tkc}. In the supernova environment, protons are non-relativistic, and the main production of dark photons is the quasi-elastic bremsstrahlung. Therefore, the flux of dark photons may be accurately calculated without introducing the virtuality form factor, see Ref.~\cite{Rrapaj:2015wgs}, and the constraint is unaffected by uncertainties. Regarding LHCb, the re-analysis would require a detailed background calculation, which is only possible using the full LHCb simulation framework and goes beyond the scope of this work.

When incorporating the uncertainties in the dark photon phenomenology, we only include those above the $\mathcal{O}(1)$ range: from the proton bremsstrahlung and the Drell-Yan process. It is meaningful because the much smaller uncertainties from the other production modes are comparable with other possible uncertainty sources that we do not include in our work (which are experiment-dependent and include the variation of the reconstruction efficiencies, effects of the finite target, etc.). Moreover, unlike the uncertainties in the bremsstrahlung and Drell-Yan process, they are invisible in the parameter space in the logarithmic scale.

\subsection{Approach to calculate the sensitivities}

To calculate the reach of various experiments in a unified fashion, we will use the \texttt{SensCalc} package~\cite{Ovchynnikov:2023cry},\footnote{May be found on Zenodo~\cite{SensCalc-Zenodo} and \href{https://github.com/maksymovchynnikov/SensCalc}{Github}} which calculates the number of events at various experiments using the semi-analytic approach. It includes all the experiments of interest and the production and decay modes of dark photons discussed in Sec.~\ref{sec:phenomenology}.\footnote{ Ref.~\cite{Foroughi-Abari:2024xlj} also presents the impact of the proton bremsstrahlung uncertainty on the reach of some experiments. However, only the future experiments SHiP and FASER2 are considered. In addition, Ref.~\cite{Foroughi-Abari:2024xlj} uses \texttt{FORESEE} package, which does not calculate the effects of the decay products acceptance -- an essential feature for beam dump experiments that significantly affect the number of events~\cite{Bondarenko:2023fex}.}

Let us briefly describe the algorithm. We begin with a detailed simulation of the production and decay of new physics particles within the decay volume. Decays of LLPs are sampled for a grid of points inside the decay volume. If the decays are exclusive, their phase space is simulated within \texttt{SensCalc}; if inclusive (i.e., includes quarks), they are showered and hadronized using \texttt{PYTHIA8} and then are used in the tabulated form. The decay products are subsequently propagated through the detector, taking into account the effects of the magnetic field on the tracker. Finally, the acceptance for decay events is determined by applying geometric criteria (such as ensuring that the decay products are within the last plane of the detector, forming a vertex within the decay volume, or maintaining a transverse impact parameter that is not too large) as well as energy and momentum requirements. 

Afterward, the simulations' output is converted into two tabulated datasets. The first one is the tabulated production distribution in LLP's mass, polar angle, and energy. The second one is the acceptance of decay events in the given experiment in the spatial coordinates and kinematics of the decaying LLP. These are then multiplied with the decay probability and integrated over LLP's decay vertices to determine the number of events. 

\texttt{SensCalc} has been tested across full-scale simulation frameworks, such as SHiP and LHCb, and toy Monte Carlo tools, including ALPINIST and FORESEE~\cite{Ovchynnikov:2023cry,Gorkavenko:2023nbk,Ovchynnikov:2022its}. In particular, using the ``Baseline'' description, its predictions reproduce public sensitivities of various experiments to dark photons, up to discrepancies caused by the different phenomenology inputs (see discussions in~\cite{Ovchynnikov:2023cry,Garcia:2024uwf}). It is currently used by the SHiP collaboration to calculate sensitivities to different long-lived particles. In the newest version, its capabilities have been extended by adding the module \texttt{EventCalc} -- the tool taking the tabulated production distributions and the decay phenomenology from \texttt{SensCalc} and sampling decay events similarly to more traditional Monte Carlo simulators. 

We implement the revised description of the production via mixing and the proton bremsstrahlung (where we include the uncertainty band), as discussed in Sec.~\ref{sec:production}; details of the implementation in \texttt{SensCalc} and \texttt{EventCalc} are given in Appendix~\ref{app:implementation-brem-SensCalc}. 

We take the description of the experiments and cuts used to calculate the acceptance for decay events from the above-mentioned collaboration papers or, if such sources are absent, from widely adopted studies performed by individuals. All the setups are implemented in \texttt{SensCalc}; the users just need to specify the selection criteria. We also refer to Ref.~\cite{Garcia:2024uwf} for further details, given that it has recently considered many of the discussed options for dark photons. We provide the main details about the setups below.

For all the experiments, we obtain the sensitivity curves by requiring $N_{\text{events}}\geq 2.3$, which corresponds to 90\% CL, assuming that they were or are expected to be background-free. We start with past and ongoing experiments and then consider future experiments. We will show their reach based on the approach of~\cite{Foroughi-Abari:2024xlj} for the bremsstrahlung and~\cite{Kyselov:2025ta} for the production via mixing. For constraints of the experiments that will not be revised in this study, we will use the curves from~\cite{Antel:2023hkf}, together with added constraints from NA62 in the dump mode from~\cite{NA62:2023qyn} and FASER~\cite{FASER:2023tle}. For the NA62 curve, the bremsstrahlung already sizeably contributes to the excluded region. Given that bremsstrahlung already contributes to the NA62 constraints and we cannot reproduce it, we have not included it.

The proton bremsstrahlung is also important for some of the constraints that are not considered in this revision, such as prompt searches at LHCb. However, to revise them, dedicated studies are needed, which we leave for future work. 

\section{Revising the reach of experiments}
\label{sec:revisiting}

\subsection{Past searches}
\label{sec:past}
\begin{figure*}[t!]
    \centering
    \includegraphics[width=0.45\textwidth]{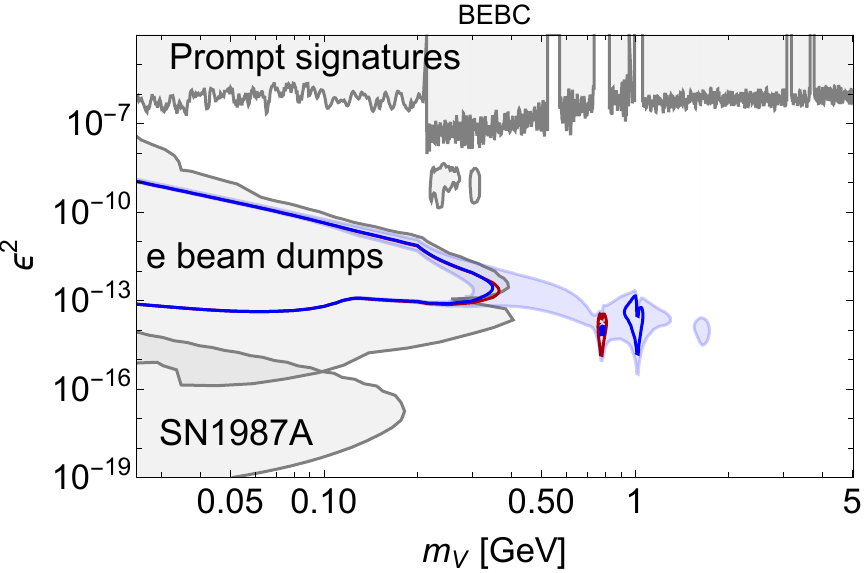}~ \includegraphics[width=0.45\textwidth]{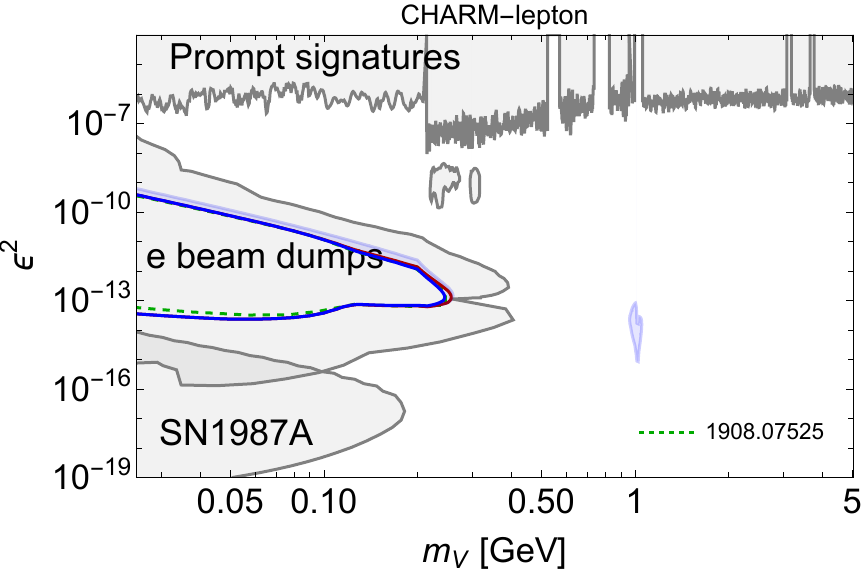} \\  \includegraphics[width=0.45\textwidth]{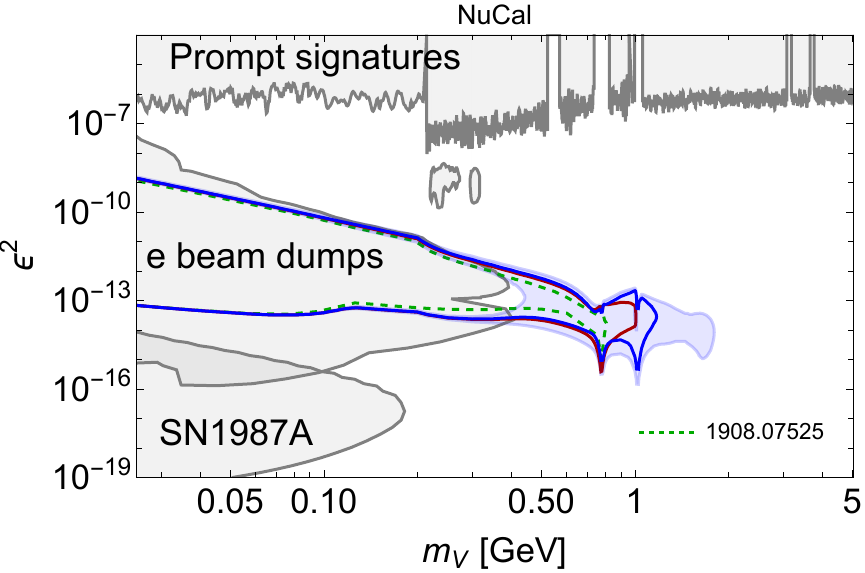}~ \includegraphics[width=0.45\textwidth]{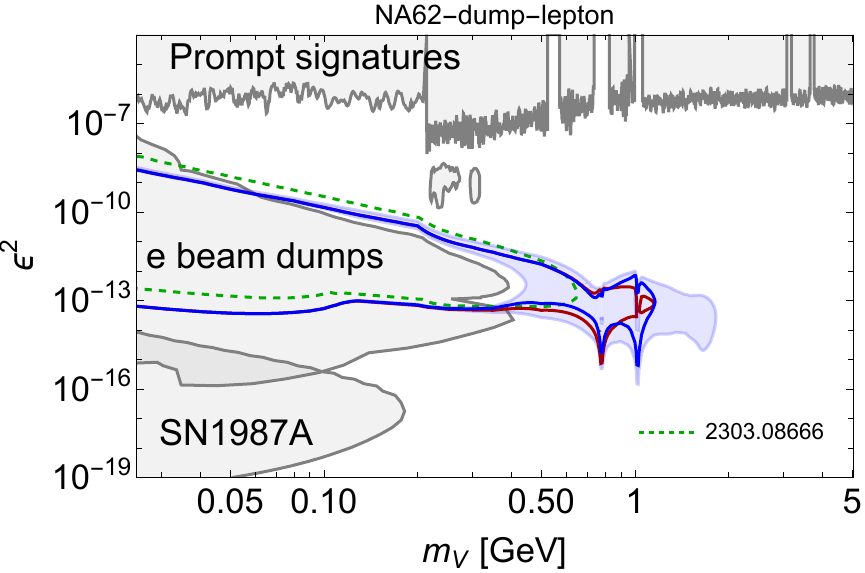}
    \caption{Parameter space of the dark photon excluded by past and ongoing experiments: BEBC (the top left panel), CHARM (the top right panel), NuCal (the bottom left panel), and NA62 in the dump mode (the bottom right panel). We consider two different descriptions of the proton bremsstrahlung, as discussed in Sec.~\ref{sec:phenomenology}: the description from~\cite{Blumlein:2013cua}, the red line, and the description from~\cite{Foroughi-Abari:2024xlj}, the blue line, with the uncertainties denoted via the light blue band. The dashed green lines show the constraints computed in past studies. The gray domain shows constraints from prompt searches, astrophysical searches, and electron and proton beam dump experiments where the proton bremsstrahlung is not involved (such as the recent search at FASER~\cite{FASER:2023tle}, which is limited by the mass range where bremsstrahlung is irrelevant). We take the corresponding curves from~\cite{Antel:2023hkf}. The isolated ``islands'' in the parameter space covered by the experiments appear because of the resonant enhancement of the dark photon production probabilities (see text for details).}
    \label{fig:past-experiments}
\end{figure*}

The past experiments of interest are BEBC, CHARM, NA62 in the dump mode, and NuCal. The parameter space probed by them as calculated within our framework is shown in Fig.~\ref{fig:past-experiments}. There, for comparison, we also include past widely adopted results. 

\begin{figure*}[t!]
    \centering
    \includegraphics[width=0.45\textwidth]{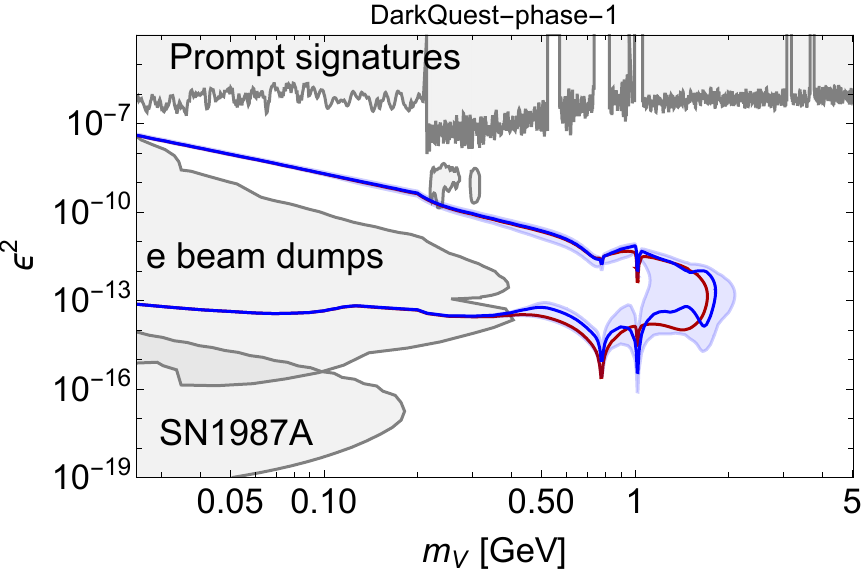}~ \includegraphics[width=0.45\textwidth]{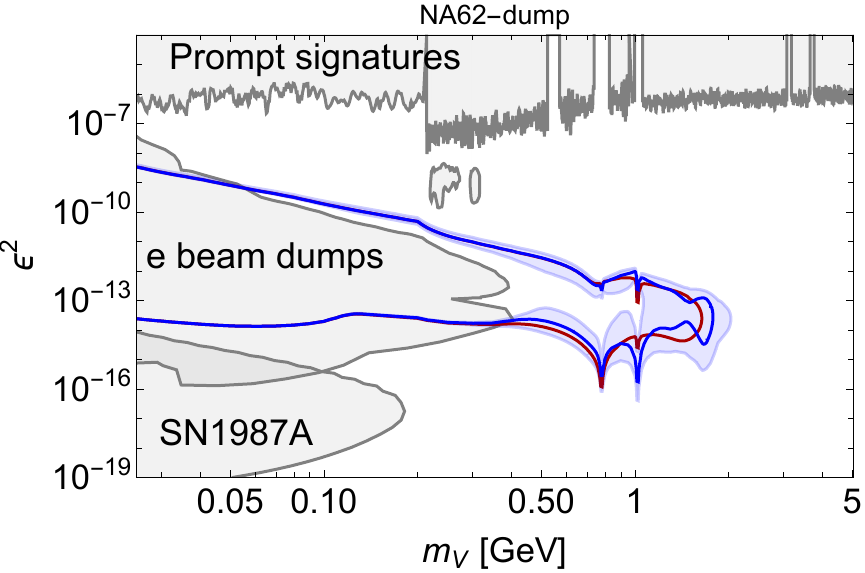} \\  \includegraphics[width=0.45\textwidth]{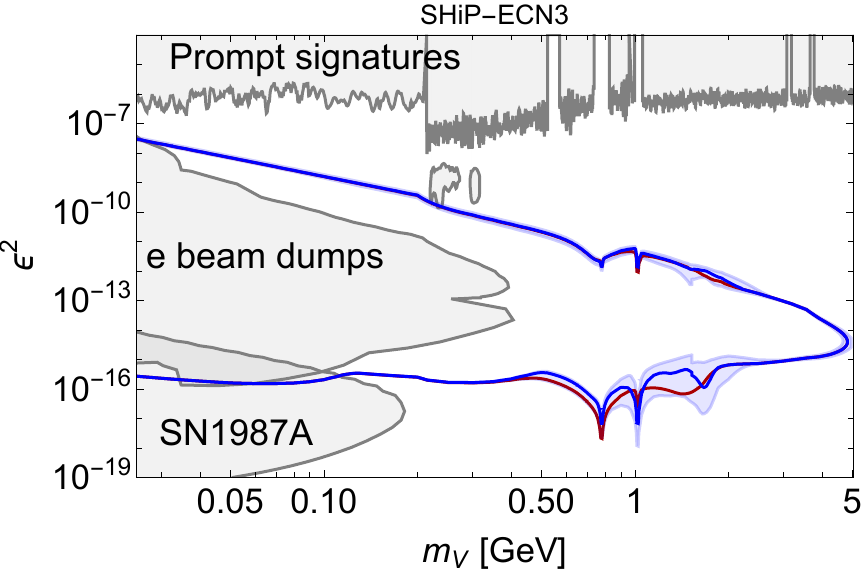}~ \includegraphics[width=0.45\textwidth]{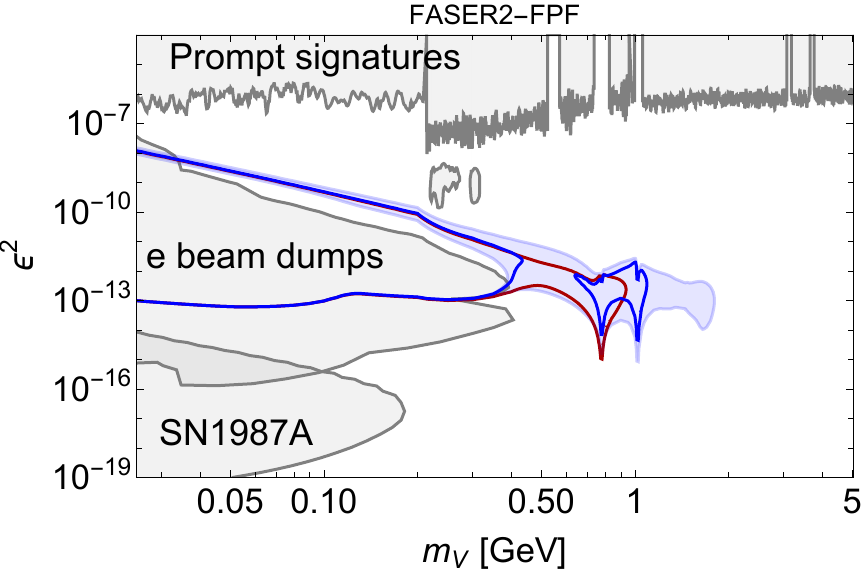}
    \caption{Parameter space of the dark photon to be probed by the future experiments: DarkQuest in phase I (the top left panel), NA62 in the dump mode (the top right panel), SHiP (the bottom left panel), and FASER2 to be located at Forward Physics Facility (the bottom right panel). Similarly to Fig.~\ref{fig:past-experiments}, the domain excluded by the past experiments does not include searches for which the proton bremsstrahlung was one of the main production channels. The meanings of the lines and gray regions are the same as in Fig.~\ref{fig:past-experiments}. For SHiP, the rapid change in the sensitivity curve at the lower bound of the bremsstrahlung's uncertainty is caused by the default inclusion of the Drell-Yan production channel starting from $m_{V} = 1.5\text{ GeV}$.}
    \label{fig:future-experiments}
\end{figure*}

Let us now discuss the results. \textit{BEBC}, similar to CHARM and NA62, operated at SPS and utilized the extracted beam of 400 GeV proton hitting the target, with the decay volume located downstream, in 404 meters. The decay volume has dimensions $3.6\times 2.5\times 1.85\text{ m}^{3}$. It effectively serves as a detector, so there is no geometric criterion on the decay products. For the latter, we consider only the electrons and muons, following the collaboration analyses and the recent re-analyses~\cite{Marocco:2020dqu,Barouki:2022bkt}, with the energy cut $E>1\text{ GeV}$. I.e., the dark photon's decay palette is restricted by the channels $V\to e^{+}e^{-}$ and $\mu^{+}\mu^{-}$, which suppresses the event yield by as large as a factor of 10, depending on the dark photon mass (see Fig.~\ref{fig:decay-br-ratios}).

The BEBC constraint on dark photons heavily varies within the uncertainty band of the proton bremsstrahlung. In particular, the maximal probed mass ranges from $\simeq 400\text{ MeV}$ up to $\simeq 1.5\text{ GeV}$, if neglecting the small isolated ``islands''. The latter appear close to the masses where the bremsstrahlung production of dark photons gets resonantly enhanced and may be explained in the following way. 

For the fixed coupling $\epsilon$, the dark photon decay length $c\tau \gamma \propto \epsilon^{-2}f(m)$ quickly decreases as a function of mass (remind Fig.~\ref{fig:decay-br-ratios}). For masses $m_{V}\gtrsim 500\text{ MeV}$, it becomes too short in the coupling range accessible at BEBC. However, in the vicinity of the SM vector resonances, the production probability gets resonantly enhanced, which opens access to lower couplings where the decay length is not so suppressed. 

The decay volume of \textit{CHARM} is located 480 m downward of the target, being a box with dimensions $3\times 3\times 35\text{ m}^{3}$. Its center is displaced relative to the beam axis by 5 m, i.e., it is an off-axis experiment. However, given the large distance to the target, the minimal angle covered by the decay volume is $\simeq 10^{-2}\text{ rad}$, which is not too large for being able to probe dark photons. The decay volume ends with a 6.4 m long detector, containing trackers and an electromagnetic calorimeter. The number of protons on target is $N_{\text{PoT}}\approx 1.7\cdot 10^{18}$. As in the collaboration analyses~\cite{CHARM:1983ayi,CHARM:1985anb} and the re-analysis~\cite{Boiarska:2021yho}, we require only leptonic decay products within the range of the calorimeter. Each of the decay products must have energy above 1 GeV. Similarly to BEBC, we only accept leptonic final states, with the reconstruction acceptance ranging between 0.65 and 0.85. 

Because of the off-axis placement, CHARM covers a smaller domain of dark photon parameter space than BEBC, despite its much larger decay volume: the dark photon flux is concentrated at tiny polar angles. The proton bremsstrahlung adds only minor modifications to the reach of CHARM, where large energy is essential to maintain sensitivity. We compare our predictions with Ref.~\cite{Tsai:2019buq}, which contains the latest calculation of the relevant constraints so far and considered the Baseline description of the proton bremsstrahlung, with somewhat different parametrization of the proton form factor. The agreement is very good, and the discrepancies come from distinct descriptions of the flux of the produced $\pi^{0}$.

\textit{NA62 in the dump mode} is characterized by a much closer placement of the decay volume compared to CHARM and BEBC -- around 102 meters, with the decay volume being approximately a cylinder with radius $1\text{ m}$, with a hole of radius of 8 cm to account for the remnant of the proton beam. It covers small polar angles $\theta< 1\text{ mrad}$. The decay volume extends in approximately $80\text{ m}$ and ends with the $\approx 63$ meters long detector system. The latter has a somewhat larger radius of $1.15\text{ m}$ of main components, including the liquid Krypton electromagnetic calorimeter (ECAL) located at its end. 

We impose the following conditions on the decay events. The decay products must remain within the detector's acceptance range until the end of the ECAL, and their momentum must exceed 1 GeV. At the beginning of the straw tracker, the minimum transverse spatial displacement between the muons must be 2 cm. For electrons or hadrons, the displacement at the entrance to the electromagnetic calorimeter must be 20 cm, ensuring that 95\% of the electromagnetic shower energy from each charged particle is spatially separated. The combination of these cuts, together with the very long detector system, significantly reduces the decay products acceptance of the NA62's detector.

To revise the constraints, we consider the setup with the number of protons on target $N_{\text{PoT}} = 1.4\cdot 10^{17}$ and only leptonic decay products (and so the states $e^{+}e^{-}$ and $\mu^{+}\mu^{-}$, remind Fig.~\ref{fig:decay-br-ratios}), following the collaboration analysis~\cite{NA62:2023qyn,NA62:2023nhs}. For comparison, we also include the bound obtained in that paper. 

Although we follow a similar description when calculating the decay rate as the collaboration paper, there is a notable difference at the lower bound in the mass domain $m_{V}\lesssim 0.1\text{ GeV}$, determined solely by the production via decays of $\pi^{0}$. In the domain at higher masses, the main reason for the difference is the description of the proton bremsstrahlung. Namely, Ref.~\cite{NA62:2023qyn} considered the description from~\cite{Blumlein:2013cua}, where the resonant enhancement of the elastic proton form factor has not been included. This setup differs even from the ``baseline'' description from Sec.~\ref{sec:phenomenology}.

Compared to the previous three experiments, \textit{NuCal} operated at a facility with a much lower proton energy $E_{p} = 70\text{ GeV}$. The lower energy would affect the overall yield of the produced dark photons and the maximal coupling to be probed (they roughly scale with $E_{p}$, see~\cite{Bondarenko:2019yob}). However, the dark photon production channels in the mass range $m\lesssim 1\text{ GeV}$ are not too sensitive to the proton beam energy. In addition, its decay volume -- a box with dimensions $2.6\times 2.6\times 23\text{ m}^{3}$ -- has been located much closer to the production point than the other experiments, at $z = 63\text{ m}$, which compensates for the smallness of $E_{p}$ at the upper bound of the sensitivity. The number of protons on target is $1.7\cdot 10^{18}$. 

We only accept the leptonic final states, with the total energy exceeding 10 GeV (here, we optimistically assume that all the energy gets deposited in the calorimeter). As in the case of CHARM, we compare our calculations with the results from~\cite{Tsai:2019buq}. The difference between the "Baseline" curves originates from the different descriptions of the proton bremsstrahlung and, similarly to the CHARM case, production of $\pi^{0}$s. 

\subsection{Future experiments}
\label{sec:future}

Let us now proceed to the currently running and future experiments. We will include NA62 in dump mode, SHiP, FASER2-FPF, and DarkQuest. Their parameter space is shown in Fig.~\ref{fig:future-experiments}. 

For \textit{NA62 in the dump mode}, we now consider $N_{\text{PoT}} = 10^{18}$ with all detectable final states, and assuming the absence of backgrounds. We use this optimistic setup to show the maximal possible reach of NA62. Under this assumption, the experiment may explore the dark photon masses as large as 1-2 GeV, depending on the bremsstrahlung flux.

\textit{SHiP} will be located at SPS. It has a large number of protons-on-target, $N_{\text{PoT}} = 6\cdot 10^{20}$. Its decay volume is an asymmetric pyramidal frustum, starting at the displacement from the target $z = 32\text{ m}$ and ending at $z = 82\text{ m}$. The transverse coverage at the largest $z$ is $\approx 4\times 6\text{ m}^{2}$. The detector system, starting at the end of the decay volume, extends by 15 m. Following~\cite{Albanese:2878604}, we require at least two decay products within the acceptance of the last sensitive plane of the detector; the particles must have a minimum momentum of $1\text{ GeV}$ and the transverse impact parameter not higher than 2.5 m.

The SHiP sensitivity extends to the masses of dark photons as large as $5\text{ GeV}$; the uncertainty caused by the proton bremsstrahlung does not affect the mass range to be probed. 

\textit{DarkQuest} is an upgrade of the SpinQuest experiment at the Fermilab beam dump facility, operating with protons having the energy $E_{p} = 120\text{ GeV}$. We will consider the phase I setup, with $10^{18}$ protons on target, effective decay volume of $2\times 2\times 5\text{ m}^{3}$ starting 7 meters downward the target, and the detector extending by 8 meters from the end of the decay volume. A possible upgrade of this experiment, phase II, will increase the number of protons by two orders of magnitude. This setup is also implemented in \texttt{SensCalc}, but we do not consider it in order not to overfill the discussion.

All the decay products are considered detectable if they are within the detector acceptance at the last plane and have energies above 1 GeV; photons must also be separated within 5.5 cm at the entrance of the ECAL (see~\cite{Blinov:2024gcw}).

In the most optimistic setup, DarkQuest's sensitivity extends to masses as large as 2 GeV. At the lower bound of uncertainty, it becomes limited to 1 GeV.

Finally, \textit{FASER2-FPF} is a recent proposal for the FASER2 experiment to be located at Forward Physics Facility~\cite{Feng:2022inv}. We consider the setup with the integrated luminosity of proton-proton collisions $\mathcal{L}=3000\text{ fb}^{-1}$ and a cylindrical decay volume with radius $R = 1\text{ m}$ and longitudinal size of $5\text{ m}$ located 620 m downward of the target. All the decay products are considered detectable if they are within the detector located downward of the decay volume. 

Similarly to the NA62 and DarkQuest cases, the bremsstrahlung uncertainty severely affects FASER2's reach. The maximal probed mass severely ranges from 0.4 to 2 GeV.

\subsection{Light dark matter}

Apart from being a standalone new physics particle, dark photons may serve as a mediator between the Standard Model and hypothetical light dark matter (LDM) $\chi$~\cite{Beacham:2019nyx}. Since the main production mechanism of $\chi$s is the decay of dark photons into $\chi \bar{\chi}$, the uncertainties in the dark photon phenomenology straightforwardly translate to the parameter space of LDM. There is a notable difference in simplifying the re-analysis: the main constraints on the $\chi$s parameter space come from the electron beam dump experiment NA64~\cite{NA64:2023wbi} and the electron-positron collider BaBar~\cite{BaBar:2017tiz}. The production modes of dark photons there are cleaner and unaffected by the uncertainties discussed in this study.

\begin{figure}[t!]
    \centering
    \includegraphics[width=\linewidth]{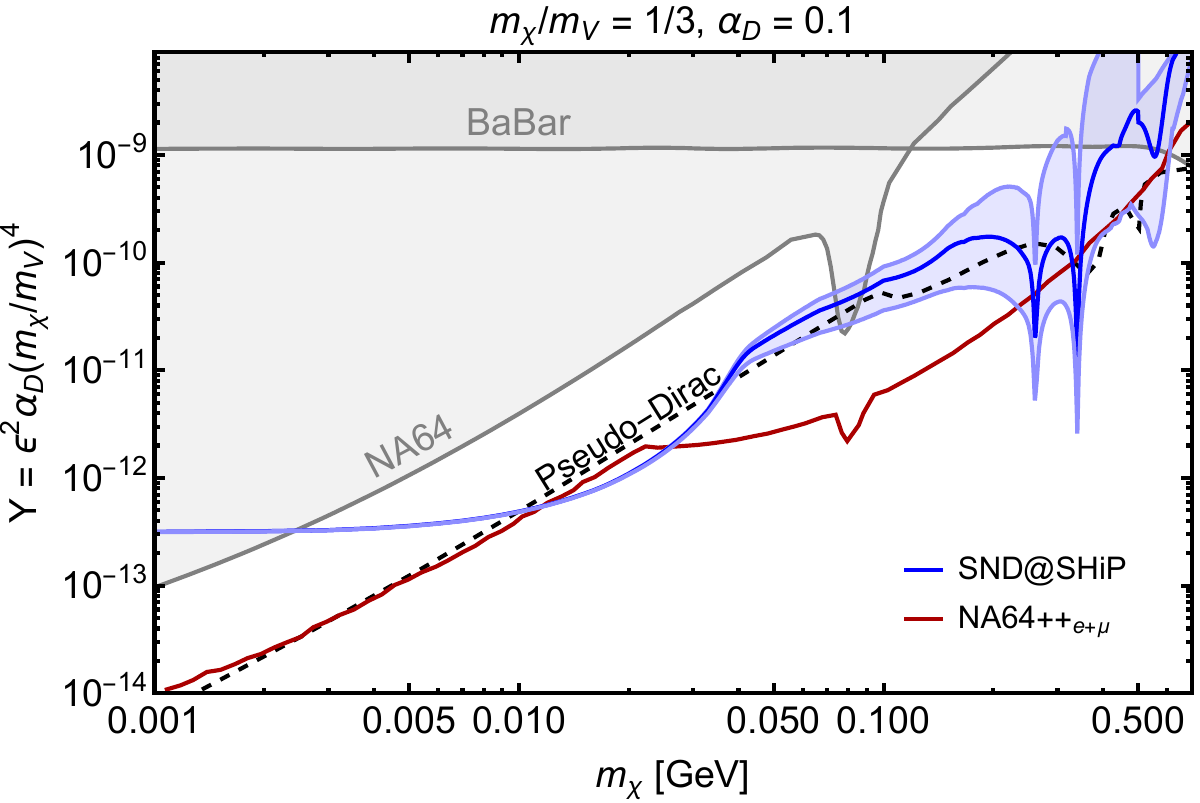}
    \caption{Sensitivity of SHiP to the model of the elastic light dark matter $\chi$ coupled via dark photons~\cite{Beacham:2019nyx,Antel:2023hkf}. Uncertainties in the SHiP sensitivity originate from the dark photon production uncertainty. The constraints are taken from~\cite{NA64:2023wbi}, the projection of the future projection of NA64 from~\cite{Gninenko:2019qiv}, and the thermal relic target line for the pseudo-Dirac $\chi$}
    \label{fig:sensitivity-ldm}
\end{figure}

The main future experiments probing the parameter space of this model are NA64~\cite{Gninenko:2019qiv} (with the increased luminosity) and SHiP~\cite{Antel:2023hkf}. SHiP will search for the events with $\chi$s via the elastic scatterings of $\chi$ off electrons in the Scattering and Neutrinos Detector (called SND@SHiP)~\cite{SHiP:2020noy}. To re-evaluate its sensitivity, we use the results of the recent study~\cite{Ferrillo:2023hhg} that performed background analysis for the new SND@SHiP setup. Instead of rescaling the sensitivity curve from~\cite{SHiP:2020noy} (as it was done in~\cite{Ferrillo:2023hhg}), we recompute the number of events from scratch.

The parameter space of LDM, including the sensitivities of SHiP and future searches at NA64, are shown in Fig.~\ref{fig:sensitivity-ldm}. The dark photon uncertainty shows up depending on the ratio between $m_{\chi}$ and $m_{V}$, which we assume to be $1/3$, following the standard convention. From the figure, we see that the uncertainty affects a huge fraction of the SHiP's target parameter space.  

\section{Conclusions}
\label{sec:conclusions}

\begin{figure}[h!]
    \centering
    \includegraphics[width=0.45\textwidth]{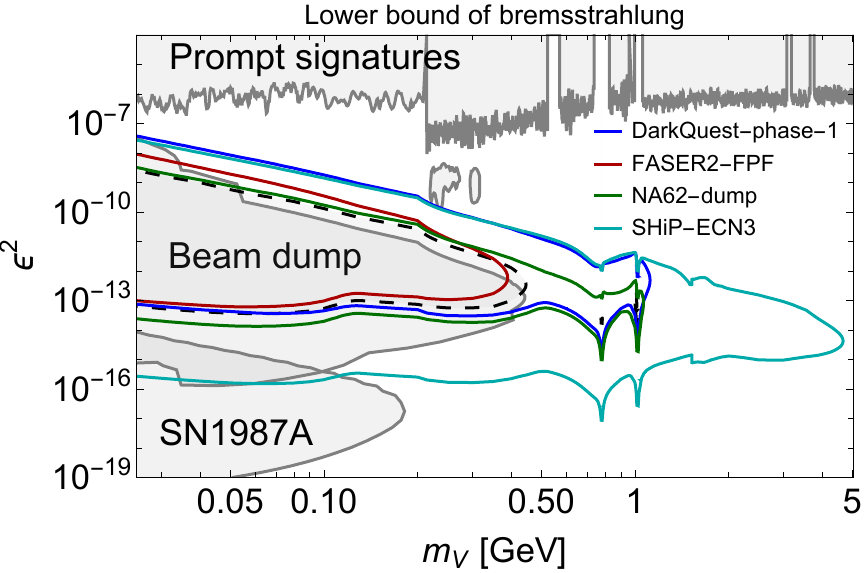} \\ \includegraphics[width=0.45\textwidth]{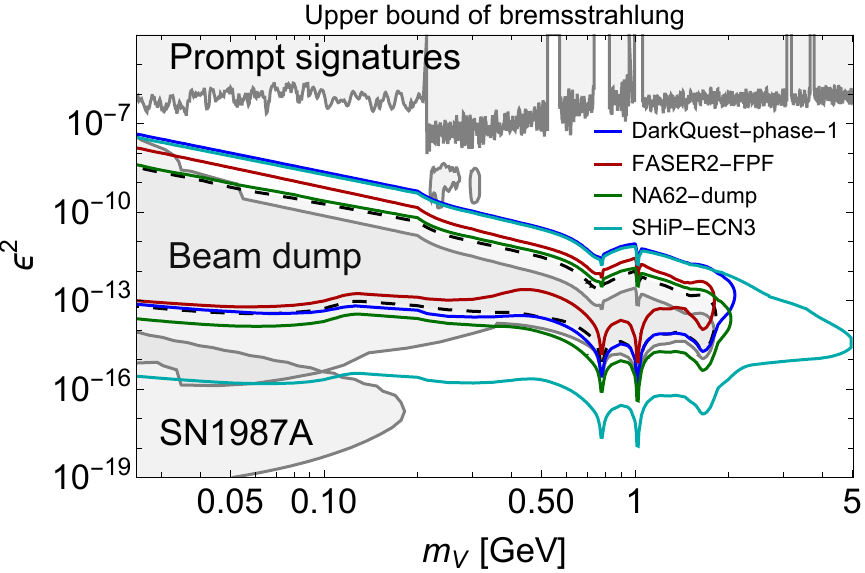}
    \caption{Summary on the dark photon parameter space. The plots show how the constraints and sensitivities to the dark photons coming from the laboratory experiments with a displaced decay volume vary depending on the description of the proton bremsstrahlung, with the upper plot showing the minimal bremsstrahlung flux and the lower corresponding to the maximal one. As for the bremsstrahlung description, we consider the approach of Ref.~\cite{Foroughi-Abari:2024xlj}, with the uncertainty band defined in Sec.~\ref{sec:phenomenology}. The gray domains show the excluded regions, including the constraints from the past experiments re-analyzed in this work (see Sec.~\ref{sec:past}). The dashed black line shows our re-analysis of the constraints from the NA62 experiment~\cite{NA62:2023nhs}. As we failed to reproduce the region from the NA62 study under a similar description of the dark photon phenomenology, Sec.~\ref{sec:past}, it will be a subject of further studies. The colored curves show the sensitivities of the future experiments (Sec.~\ref{sec:future}).}
    \label{fig:final}
\end{figure}

In this work, we have presented a comprehensive and systematic analysis of the potential reach of various proton accelerator experiments in detecting long-lived dark photons. 

We have addressed two issues persisting in the community. First, we have used the latest developments in dark photon phenomenology. They include production mechanisms such as proton bremsstrahlung (including the theoretical uncertainty) and mixing with neutral mesons, as well as decay channels into various hadrons, see Sec.~\ref{sec:phenomenology}. This has allowed us to properly define the variation of the constraints and sensitivities of different experiments depending on the uncertainty in the production modes, as well as accurately calculate the acceptance for their decay products.

Second, we have systematically incorporated phenomenology in the \texttt{SensCalc} tool, which calculates the event rate with LLPs for the broad range of experiments at different facilities, see Sec.~\ref{sec:method}. This has provided consistency in calculations of experimental reach -- the feature that has been missing in the literature.

Using this machinery, we have carefully analyzed how the constraints and sensitivities of various experiments change depending on the variation of the proton bremsstrahlung flux -- the main source of uncertainty in the event rate (Sec.~\ref{sec:phenomenology}), which may change it by orders of magnitude, see Sec.~\ref{sec:revisiting}. We have found that this translates into varying the maximal mass probed by the given experiment by as large as a factor of two, depending on the setup, both for past (Sec.~\ref{sec:past}) and future (Sec.~\ref{sec:future}) experiments. 

To demonstrate the magnitude of the uncertainties, Fig.~\ref{fig:final} shows the parameter space of dark photons where the proton bremsstrahlung flux is considered at its lower and upper values within the uncertainty band. 

Our findings are significant for several reasons. First, recognizing that the constraints from past experiments are heavily influenced by theoretical uncertainties reshapes the target parameter space for future searches. Second, as the sensitivity of upcoming experiments, such as FASER2, varies considerably, their physics case and overall appeal require careful reassessment. Third, we have identified a critical limitation in the state-of-the-art studies~\cite{Beacham:2019nyx,Antel:2023hkf}: the lack of unified recommendations for the experimental community regarding the phenomenology of long-lived particles. This inconsistency leads to incoherent presentations of constraints and sensitivities across various experiments in summary figures. Addressing this issue promptly is essential to fully assess the potential of future experiments to explore the parameter space.

In the end, let us highlight the directions to be explored. First, the revision of the dark photon parameter space performed in this work is far from being complete. Constraints and sensitivities from the other probes, such as prompt searches at the LHC, may rely on the dark photon bremsstrahlung and thus have to be recomputed. 

Dark photons are not the only well-motivated model of LLPs where the theoretical uncertainties in the production and decay modes may severely affect the status of the parameter space. The impact of the theoretical uncertainty in the phenomenology on the probed parameter space for Higgs-like scalars, $B-L$ mediators, axion-like particles, and other LLPs has to be carefully analyzed. However, these models also have uncertainties coming from the production via the proton bremsstrahlung and mixing with neutral mesons~\cite{Monin:2018lee,Blackstone:2024ouf,Smith:2024jve}. Therefore, these studies are a subject of dedicated research. This paper indicates the unified and systematic approach to how this may be done. We leave these interesting questions for future work. 

\textbf{Acknowledgements.} The authors thank Adam Ritz and Saeid Foroughi-Abari for providing valuable comments on the proton bremsstrahlung production channel and the manuscript.

\onecolumngrid 

\appendix

\section{Production of dark photons in the bremsstrahlung process}
\label{app:bremsstrahlung}

The proton bremsstrahlung is, in general, an inelastic process: although the $A$ gets emitted from the elastic $ppA$ vertex, the final hadronic state originates from inelastic collisions of protons. The exact description of such a process does not exist in the literature. However, under certain kinematic conditions, its differential cross-section may be represented as a product of the proton-proton inelastic cross-section $\sigma_{pp\to X}$, which we know from the experimental observations, times the so-called splitting probability $\omega_{\text{spl}}$ describing the probability of emitting the dark photon by an incoming proton:
\begin{equation}
    \frac{d\sigma_{pp \to A+X}^{\text{brem}}}{dp_{T}^{2}dz} \approx \omega_{\text{spl}}(p_{T},z)\times \sigma^{\text{ISR}}_{pp\to X}(s_{pp}(p_{T},z)),
\label{eq:splitting-probability}
\end{equation}
where $p_{T},z$ are dark photon's transverse momentum and momentum fraction $z = p_{A}/p_{p}$ respectively, and $\sigma^{\text{ISR}}$ is the total initial state radiation cross-section, which may be approximated by the non-single-diffractive (NSD) cross-section~\cite{Foroughi-Abari:2021zbm}.

\begin{figure*}[h!]
    \centering
    \includegraphics[width=\textwidth]{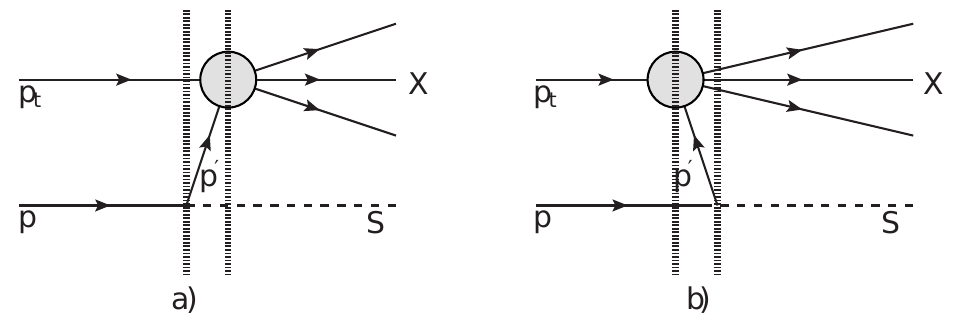}
    \caption{Advanced (a) and retarded (b) diagrams describing the production of a particle $S$ in the proton bremsstrahlung process within the framework of old-fashioned perturbation theory. The figure is taken from Ref.~\cite{Boiarska:2019jym}.}
    \label{fig:brem-diagrams}
\end{figure*}

The idea of the approximation may be transparently illustrated by considering the so-called old-fashioned perturbation theory, which considers time as a dedicated coordinate and finds an approximate solution of Schroedinger's equation~\cite{Altarelli:1977zs}. Under this description, there are two diagrams for the bremsstrahlung process: the so-called advanced and retarded one, see Fig.~\ref{fig:brem-diagrams}. The description below closely follows Ref.~\cite{Boiarska:2019jym}, which applied the approach to the Higgs-like scalars. The corresponding matrix elements $V_{a},V_{r}$ have the form
\begin{equation}
    \mathcal{V}_{a} = \frac{\mathcal{M}_{p\to p'A}\mathcal{M}_{p'p_{t}\to X}}{2E_{p'}(E_{p'}+E_{A}-E_{p})}\bigg|_{\bm{p}_{p'} = \bm{p}_{p}-\bm{p}_{A}}, \quad \mathcal{V}_{r} = \frac{\mathcal{M}_{p_{t}\to p'X}\mathcal{M}_{p'p\to S}}{2E_{p'}(E_{p}+E_{p'}-E_{A})}\bigg|_{\bm{p}_{p'} = \bm{p}_{p}-\bm{p}_{A}}
    \label{eq:matrix-element-old-fashioned}
\end{equation}
Here, $\mathcal{M}$ denotes the Lorentz-invariant amplitude of the processes, $p$ is the incoming proton which emits $A$, $p'$ is the intermediate proton, while $p_{t}$ stays for the other proton which is collided with $p$. The total matrix element, $\mathcal{V} = \mathcal{V}_{a}+\mathcal{V}_{r}$, may lead to the cross-section~\eqref{eq:splitting-probability} if $V_{r}$ can be neglected compared to the advanced matrix element $V_{a}$. This is the case if the following condition is satisfied:
\begin{equation}
    \frac{E_{p'}+E_{S}-E_{p}}{E_{p}+E_{p'}-E_{S}} \approx \frac{p_{T}^{2}+(1-z)m_{S}^{2}+z^{2}m_{p}^{2}}{4p_{p}^{2}z(1-z)^{2}} \ll 1,
    \label{eq:WW-approximation}
\end{equation}
The last step is to consider the two differential cross sections -- one for the full process $pp_{t}\to A+X$, and another one for the sub-process $p'p_{t}\to X$:
\begin{align}
    d\sigma_{pp_{t}\to AX} =& \frac{1}{4E_{p}E_{p_{t}}}\frac{|\mathcal{M}_{p\to p'A}|^{2}|\mathcal{M}_{p'p_{t}\to X}|^{2}}{(2E_{p'})^{2}(E_{p'}+E_{A}-E_{p})^{2}}(2\pi)^{4}\delta^{(4)}\left(p_{p}+p_{p_{t}}-p_{A}-\sum_{X}p_{X}\right) \frac{d^{3}\bm{p}_{A}}{(2\pi)^{3}2E_{A}} \prod_{X}\frac{d^{3}\bm{p}_{X}}{(2\pi)^{3}2E_{X}},
    \label{eq:sub-process-a}
    \\
    d\sigma_{p'p_{t}\to X} =& \frac{1}{4E_{p'}E_{p_{t}}}|\mathcal{M}_{p'p_{t}\to X}|^{2} (2\pi)^{4}\delta^{(4)}\left(p_{p'}+p_{p_{t}}-\sum_{X}p_{X}\right) \prod_{X}\frac{d^{3}\bm{p}_{X}}{(2\pi)^{3}2E_{X}}
    \label{eq:sub-process-b}
\end{align}
Neglecting the difference in the energy conservation arguments in the delta-functions that are of order $\mathcal{O}(m_{p/S}^{2}/p_{p}^{2},p_{T}^{2}/p_{p}^{2})$, we can relate these cross-sections via Eq.~\eqref{eq:splitting-probability}.

From its definition, the splitting probability $\omega_{\text{spl}}$ has to include two form factors staying for distinct reasons. The first one is the proton elastic form factor $F_{T}$ in the timeline region of the transferred momentum $t = (p-p')^{2} = m_{V}^{2}$, accounting for resonant contributions from intermediate mesons $\rho^{0},\omega,\phi$ and their excitations. The second one is the ``virtuality'' form 
factor $\mathcal{F}_{V}$ ensuring the smallness of the departure of $p'$ from the real proton. 

Ref.~\cite{Foroughi-Abari:2021zbm} used this approach (which we will call the Altarelli-Parisi approach, or AP, by the names of the authors of Ref.~\cite{Altarelli:1977zs}) to calculate the splitting probability for dark photons:
\begin{equation}
\omega_{\text{spl}}= \frac{\alpha_{\text{EM}}\epsilon^{2}}{2\pi}|F_{T}|^{2}\mathcal{F}_{V}^{2}\times \\ \frac{1}{H}\left(z-z(1-z)\frac{2m_{p}^{2}+m_{A}^{2}}{H}+\frac{H}{2zm_{A}^{2}} \right),
    \label{eq:splitting-probability-AP}
\end{equation}
Here, $H(z,p_{T}) = p_{T}^{2}+z^{2}m_{p}^{2}+(1-z)m_{A}^{2}$. $F_{T}$ is given by the sum of Breit-Wigner contributions from $\rho^{0},\omega$ and their two lower excitations 
\begin{equation}
    F_{T} \approx \sum_{m^{0} = \rho,\omega}\frac{f_{m^{0}}m_{m^{0}}^{2}}{m_{m^{0}}^{2}-m_{A}^{2}-im_{m^{0}}\Gamma_{m_{m^{0}}}},
    \label{eq:form-factor-timelike}
\end{equation}
where the couplings $f_{m^{0}}$ are fixed either from the data or from the requirement for $F_{T}$ to vanish in the limit $m_{A}\gg m_{m^{0}}$ following by quark counting rule. 

There is no unique way to fix the virtuality form factor. Therefore, Ref.~\cite{Foroughi-Abari:2021zbm} considered the phenomenological form factor
\begin{equation}
\mathcal{F}_{V} = \frac{1}{1+\left(\frac{p'^{2}-m_{p}^{2}}{\Lambda_{p}^{2}}\right)^{2}}\approx \frac{1}{1+\frac{H^{2}}{z^{2}\Lambda_{p}^{4}}},
\label{eq:reality-condition}
\end{equation}
where $\Lambda_{p}$ denotes a cut-off scale. Its value should not significantly exceed $m_{p}$.

Ref.~\cite{Blumlein:2013cua} used another approach in deriving the probability of the proton bremsstrahlung -- considering the quasi-elastic scattering of a proton off another proton via emission of the dark photon and extrapolating it to the full inelastic cross-section. The quasi-elastic scattering is described formally by exchanging a vector particle $q$ between the protons and considering the Weiszacker-Williams approximation, i.e., approximating the $2\to 3$ quasi-elastic process by the $2\to 2$ process with ``almost real'' $q$ times the splitting probability. This approach has been then ``extrapolated'' onto full inelastic scattering. The splitting probability takes the form 
\begin{equation}
\omega_{\text{spl}} = \frac{\alpha_{\text{EM}}\epsilon^{2}}{2\pi}|F_{T}|^{2}\frac{1}{H}\bigg[\frac{1+(1-z)^{2}}{z} -2z(1-z)\bigg\{ \frac{2m_{p}^{2}+m_{A}^{2}}{H}-z^{2}\frac{2m_{p}^{4}}{H^{2}}-(1+(1-z)^{2})\frac{m_{p}^{2}m_{A}^{2}}{H^{2}}-(1-z)\frac{m_{A}^{4}}{H^{2}}\bigg\} \bigg]
\label{eq:splitting-probability-Baseline}
\end{equation}
The transverse momentum cut $p_{T}<1\text{ GeV}$ and the limitation $0.1<z<0.9$ were used, corresponding to the angular coverage of the NuCal experiment for which the study~\cite{Blumlein:2013cua} has been performed. The approach~\eqref{eq:splitting-probability-Baseline} with similar cuts is widely used by the scientific community to study the parameter space of dark photons excluded by past experiments and to be probed by future ones~\cite{Beacham:2019nyx,Berlin:2018jbm,Tsai:2019buq,SHiP:2020vbd,SHiP:2020sos,Kling:2021fwx,Antel:2023hkf,Ovchynnikov:2023cry}, although utilizing slightly various expressions for the timelike form-factor.

There are two differences in the splitting probabilities~\eqref{eq:splitting-probability-AP},~\eqref{eq:splitting-probability-Baseline}. First, the probability~\eqref{eq:splitting-probability-AP} contains an explicit divergence in the limit $m_{A}\to 0$, which is similar to the unphysical soft photon bremsstrahlung singularity. In contrast, the expression~\eqref{eq:splitting-probability-Baseline} stays finite. The second one is treating the virtuality of the intermediate proton. In the approach~\eqref{eq:splitting-probability-AP}, there is a clear definition of the splitting probability using the two processes~\eqref{eq:sub-process-a},~\eqref{eq:sub-process-b}. The virtuality form factor~\eqref{eq:reality-condition} regularizes this relation, although leaving room for the variation of the regularization scale $\Lambda_{p}$. On the contrary, the approach~\eqref{eq:splitting-probability-Baseline} generalizes the quasi-elastic splitting probability onto the full inelastic process. This generalization does not provide a clear recipe on how to introduce the virtuality regulator, although it is clear that it should be restricted.

To fix the issue with the infrared singularity, the recent study~\cite{Foroughi-Abari:2024xlj} utilized the effective approach of~\cite{Dawson:1984gx} to remove unphysical contributions from the longitudinal polarizations of $V$ to the splitting probability in the quasi-real approximation~\eqref{eq:splitting-probability-AP}. It also accounted for the dipole form factor in the effective expression of the nucleon matrix element of the EM current 
\begin{equation}
\langle N |J_{\text{EM},\mu} |N\rangle = \bar{u}(p')\left(\gamma_{\mu}F^{N}_{1}(t) + \frac{i\sigma_{\mu\nu}(p-p')^{\nu}}{2m_{N}} F_{2}^{N}(t)\right)u(p), \quad t = (p-p')^{2}
\end{equation}
where $N = p/n$, and $F_{1},F_{2}$ are the Dirac and Pauli form factors in the timelike region. The resulting $\omega_{\text{spl}}$ is
\begin{equation}
\omega_{\text{spl}} = \mathcal{F}_{V}^{2}\cdot \bigg(w_1^{\text{eff}}(z, p_T^2) |F_1(m_{V})|^2 + w_2^{\text{eff}}(z, p_T^2) |F_2(m_{V})|^2 +w_{12}^{\text{eff}}(z, p_T^2) \, \text{Re} \left[ F_1(m_{V}) F_2^*(m_{V}) \right]\bigg),
\label{eq:splitting-probability-FR}
\end{equation}
where the coefficients $\omega^{\text{eff}}_{1},\omega^{\text{eff}}_{2},\omega^{\text{eff}}_{3}$ are
\begin{align}
    w_1^{\text{eff}}(z, p_T^2) =&\frac{\alpha_{\text{EM}} \epsilon^2}{2 \pi H} \bigg[ \frac{(1 + (1 - z)^2)}{z} - z(1 - z) \left( \frac{2m_p^2 + m_V^2}{H} \right) \bigg],
\\ 
    w_2^{\text{eff}}(z, p_T^2) =& w_1^{\text{eff}}(z, p_T^2)  +\frac{\alpha_{\text{EM}} \epsilon^2}{4 \pi m_p^2} \left[ 1 - (1 - z) \left( \frac{4m_p^2 + m_V^2}{H} \right) \right] \times \left[ \frac{1}{z} - \frac{1}{4}z \left( \frac{4m_p^2 - m_V^2}{H} \right) \right],
\\ 
w_{12}^{\text{eff}}(z, p_T^2) =& \frac{\alpha_{\text{EM}} \epsilon^2}{2 \pi H} \left[ 2z - z(1 - z) \frac{3m_V^2}{H} \right]
\end{align}
$\omega^{\text{eff}}_{1}$ matches the splitting probability obtained using the quasi-real approximation, whereas the remaining two terms originate from the presence of the dipole vertex parametrized by the form factor $F_{2}$.

Compared to the previous studies as outlined by Eq.~\eqref{eq:form-factor-timelike}, contributions from $\phi$ meson and its excited states have been included in $F_{1,2}$. To estimate the theoretical uncertainty, they have been calculated using two different approaches -- dispersion relations method and unitarity and analytic model~\cite{Dubnicka:2002yp}, which requires $F_{1,2}$ to be analytic functions throughout the entire complex plane of the transferred momentum, except for cuts extending along the positive real axis from the lowest continuum branch point $t_{0}$ to infinity.

\section{Implementation of new production description in \texttt{SensCalc} and \texttt{EventCalc}}
\label{app:implementation-brem-SensCalc}

Various implementations of the description of the dark photon production channels are implemented in the newest version of \texttt{SensCalc}. 

Let us first consider the proton bremsstrahlung. The descriptions are different in terms of the splitting probability and the model for the elastic proton form factor. They are available through the notebook \texttt{2. LLP distribution.nb} and the module \texttt{bremsstrahlung.nb} located in the folder \textit{codes}, that is uploaded by the notebook. The notebook produces tabulated angle-energy distributions of various LLPs, including dark photons, which are then used by the other notebooks to produce sensitivities. 

Details are given in Table~\ref{tab:brem-descriptions-SensCalc}.

\begin{table}[h!]
    \centering
    \begin{tabular}{|c|c|c|c|c|c|}
    \hline Name in \texttt{SensCalc} & $\omega_{\text{spl}}$, ref. & $F_{1,2}$, Ref. & Varied parameters & Uncertainty \\ \hline 
    "Baseline"  & \cite{Blumlein:2013cua} & \cite{Berlin:2018jbm} & $p_{T,\text{max}},z_{\text{min}/\text{max}}$ & -- \\ \hline 
      "AP" & \cite{Foroughi-Abari:2021zbm} & \cite{Berlin:2018jbm} & $z_{\text{min}/\text{max}},\Lambda_{p}$ & $\Lambda_{p} \in (0.5,2)\text{ GeV}$ \\ \hline 
       "FR" & \cite{Foroughi-Abari:2024xlj} & \cite{Foroughi-Abari:2024xlj} & \makecell{$z_{\text{min}/\text{max}},\Lambda_{p}$\\ Form-factor $ppV$} & \makecell{$\Lambda_{p} \in (0.5,2)\text{ GeV}$ \\ Variation of the form-factor \\ according to the UA model} \\ \hline 
    \end{tabular}
    \caption{Various descriptions of the production of dark photons via bremsstrahlung as implemented in the newest version of \texttt{SensCalc}: The name as implemented, the reference for the splitting probability used in the description, the reference for the elastic proton-proton-dark photon form factor, the list of varied parameters, and the definition of the uncertainty within the approach. The "Baseline" option corresponds to the baseline description widely used by the community; in particular, it was incorporated by the event generators \texttt{FORESEE}~\cite{Kling:2021fwx} and the previous versions of \texttt{SensCalc}. The "FR" approach follows Ref.~\cite{Foroughi-Abari:2024xlj}. The "AP" description is the intermediate setup described~\cite{Foroughi-Abari:2021zbm}. Although it contains the infrared singularity, it has an advantage over the ``Baseline'' option as it has defined uncertainty. It may be applied for the other particles, such as dark scalars (see also~\cite{Boiarska:2019jym,Batell:2020vqn}, for which the bremsstrahlung studies are currently in a less advanced stage.}
    \label{tab:brem-descriptions-SensCalc}
\end{table}

By default, in the dark photon case, \texttt{SensCalc} exports the tabulated distributions for ``Baseline'' and ``FR''. For ``Baseline'', it assumes $p_{T,\text{max}} = 1\text{ GeV}$ independently of the facility (it may be changed in the \texttt{bremsstrahlung.nb}). For \texttt{FR}, it assumes the variation of the proton elastic form factor according to the UA model (remind Fig. 3 from~\cite{Foroughi-Abari:2024xlj}), as well as the variation of the virtuality scale $\Lambda_{p}$ in the range $0.5-2\text{ GeV}$ (this range may be changed in the code block \texttt{BlockTabulatedPDFsFromBrem[LLP\_, Facility\_, Description\_]} of the \texttt{2. LLP distribution.nb}). 

Let us now discuss the production via mixing. The description used previously in \texttt{SensCalc} is now called ``Mixing-Old''. For the production cross-section, it takes the product of the $\rho^{0}$ cross-section times the squared mixing angle between the dark photons and $\rho$s (following Appendix D of \cite{Berlin:2018jbm}), while for the kinematics it takes the description based on~\cite{Jerhot:2022chi}. The corresponding tabulated probabilities The revised description is called ``Mixing'' and is available in the form of precomputed angle-energy distributions stored in the folder \textit{spectra/New physics particles spectra/DP/Pregenerated}.

These descriptions are then used by the notebook \texttt{3. DP sensitivity.nb}. Once launching, users may select between various descriptions of the production via mixing. Regarding the bremsstrahlung, all the generated descriptions are involved to demonstrate the variation of the bremsstrahlung flux due to the variation of the ``FR'' setup, and how it compares with the previous state-of-the-art approach. Once the tabulated event rate for different production channels is computed, the notebook calculates the sensitivity curve by switching between different bremsstrahlung descriptions. It then exports the association between the bremsstrahlung description and the calculated sensitivities in \texttt{.mx} and \texttt{.json} files. 

Finally, the generated sensitivities are used for making plots in the notebook \texttt{4. Plots.nb}. It utilizes the joined constraints on dark photons depending on the bremsstrahlung description, including also the option without including the bremsstrahlung. The corresponding constraints may be found in the .mx and .json files located in the `contours/DP` directory.

Using the tabulated distributions produced by \texttt{SensCalc} as well as the information about the production and decay channels, \texttt{EventCalc} samples events with decaying dark photons. Similarly to the case of \texttt{SensCalc}, users are asked to choose the description of the proton bremsstrahlung and the mixing with neutral mesons. More details on \texttt{EventCalc} is provided in the updated manual of the \texttt{SensCalc} repository, which may be found on~\href{https://zenodo.org/doi/10.5281/zenodo.7957784}{Zenodo} and \href{https://github.com/maksymovchynnikov/SensCalc}{GitHub}.

\bibliography{bib.bib}

\end{document}